\documentclass{article}
\usepackage[utf8]{inputenc}
\usepackage{xcolor}
\usepackage{authblk}

\title{Planetary Migration } 
\author[1]{J.C.B. Papaloizou}
\affil[1]{DAMTP, University of Cambridge,  Wilberforce Road, Cambridge CB3 0WA, UK}
\date{}
\usepackage{natbib}
\usepackage{graphicx}
\usepackage{geometry}
\usepackage[utf8]{inputenc}
\usepackage{wasysym}

\begin{document}
\maketitle
\section{Introduction} 
Studies of planet migration derived from disc planet interactions  began before the discovery of exoplanets \citep[see][for reviews]{Petal2007,Betal2014}.
The  potential importance of migration for determining orbital architectures being realised,  the field received greater attention soon after the initial discoveries of exoplanets.
Early studies based on very simple disc models indicated very fast migration times for low mass planets that raised questions about  its relevance.
However, more recent studies, made possible with improving  resources, that considered improved physics and disc models revealed
processes that could halt or reverse this migration. That in turn led to a focus on special  regions in the.disc where migration could be halted.
In this way the migration of low mass planets could be reconciled with formation theories.
In the case of giant planets which have a nonlinear interaction with the disc, the migration should be slower and coupled to the evolution of the disc
which is where attention needs to be focused. 

This review is primarily concerned with  processes where migration is connected with the presence of
the protoplanetary disk as described above.
 However, it should be noted that migration  may be induced by disc-free gravitational interactions amongst planets or with binary companions.
This  will be  discussed relatively briefly.

\section{State of the art opportunities and challenges}  
The classification of the main types of migration applicable to single planets: type I, type II and type III which apply for different planet masses and disc physics
has been established. In addition systems of planets have been considered in order to investigate the origin and sustainability of commensurabilities.
 However, the disc models normally considered are primitive in comparison to those that are becoming available which involve magnetic fields, detailed vertical 
 structure and dust.  Access to steadily improving computer resources and numerical tools should enable studies of disc planet interactions in their context.
 We may then obtain a more complete picture of how and  where migration stalls for low mass planets and the speed of migration of giant planets.
 After brief descriptions of the main types of migration  in Sections \ref{Gen}-\ref{Type3}  we give an outline of some of the issues
that may be addressed with improved resources in future in Sections \ref{goals}-\ref{Multi}.

\begin{figure*}
\vspace{-10cm}
\hspace{0.0cm} \includegraphics[trim=1 -85 1 300,clip,width=0.4\textwidth]{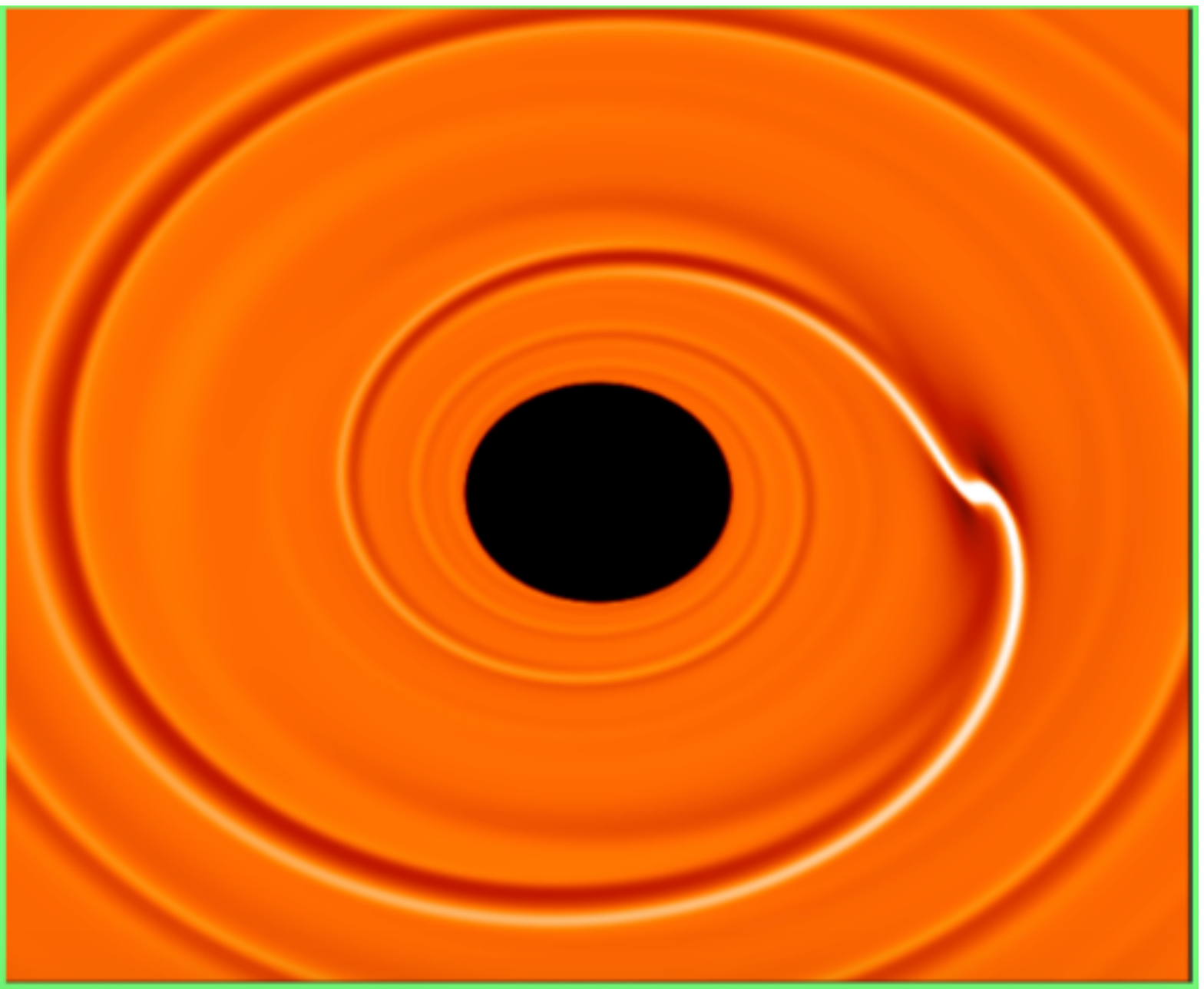}
\hspace{-0.5cm} \includegraphics[trim=1 4 1 10,clip,width=0.27\textwidth]{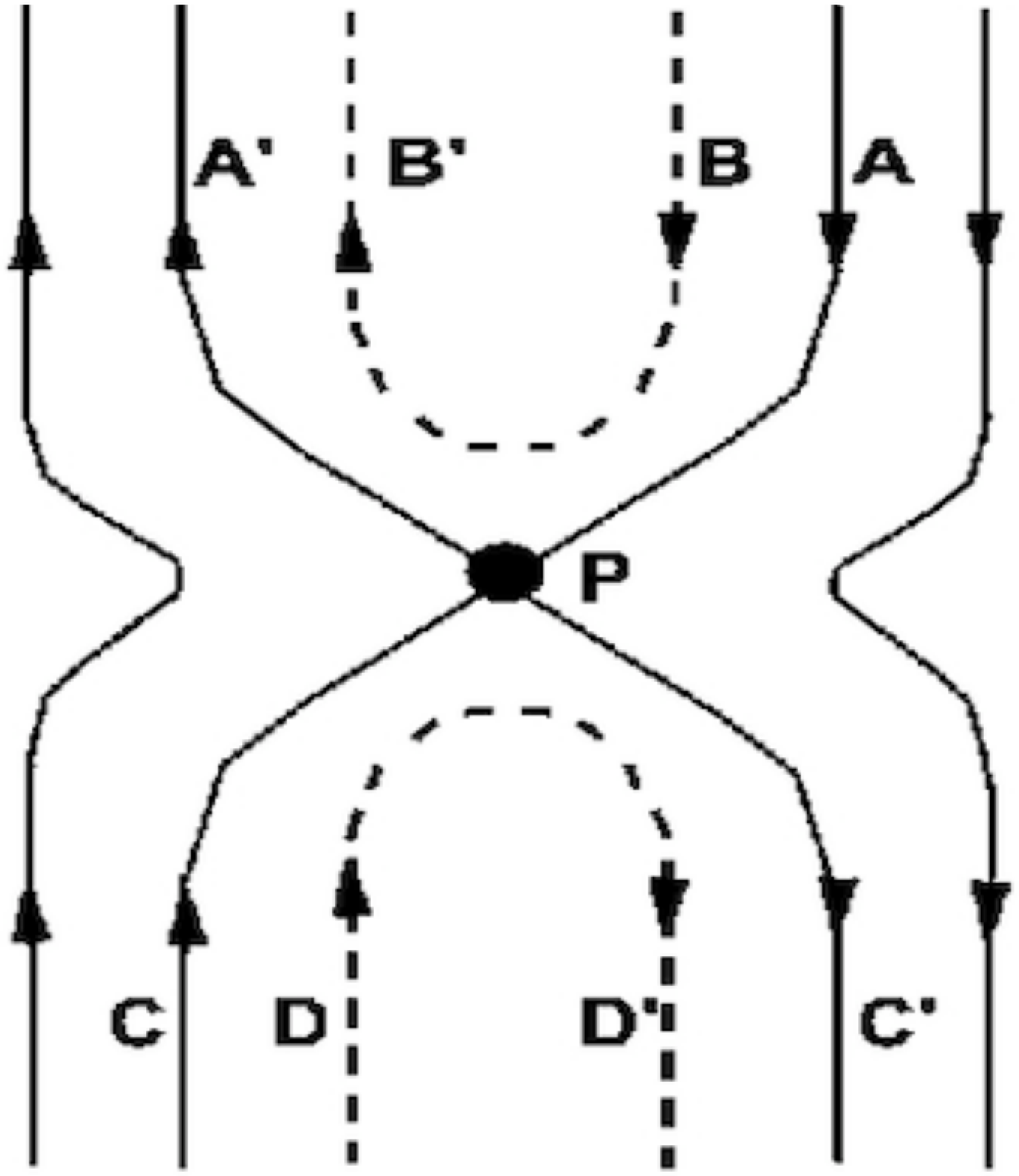}
\hspace{-5.2cm} \includegraphics[trim=1 320  0 15,clip,width=1.1\textwidth]{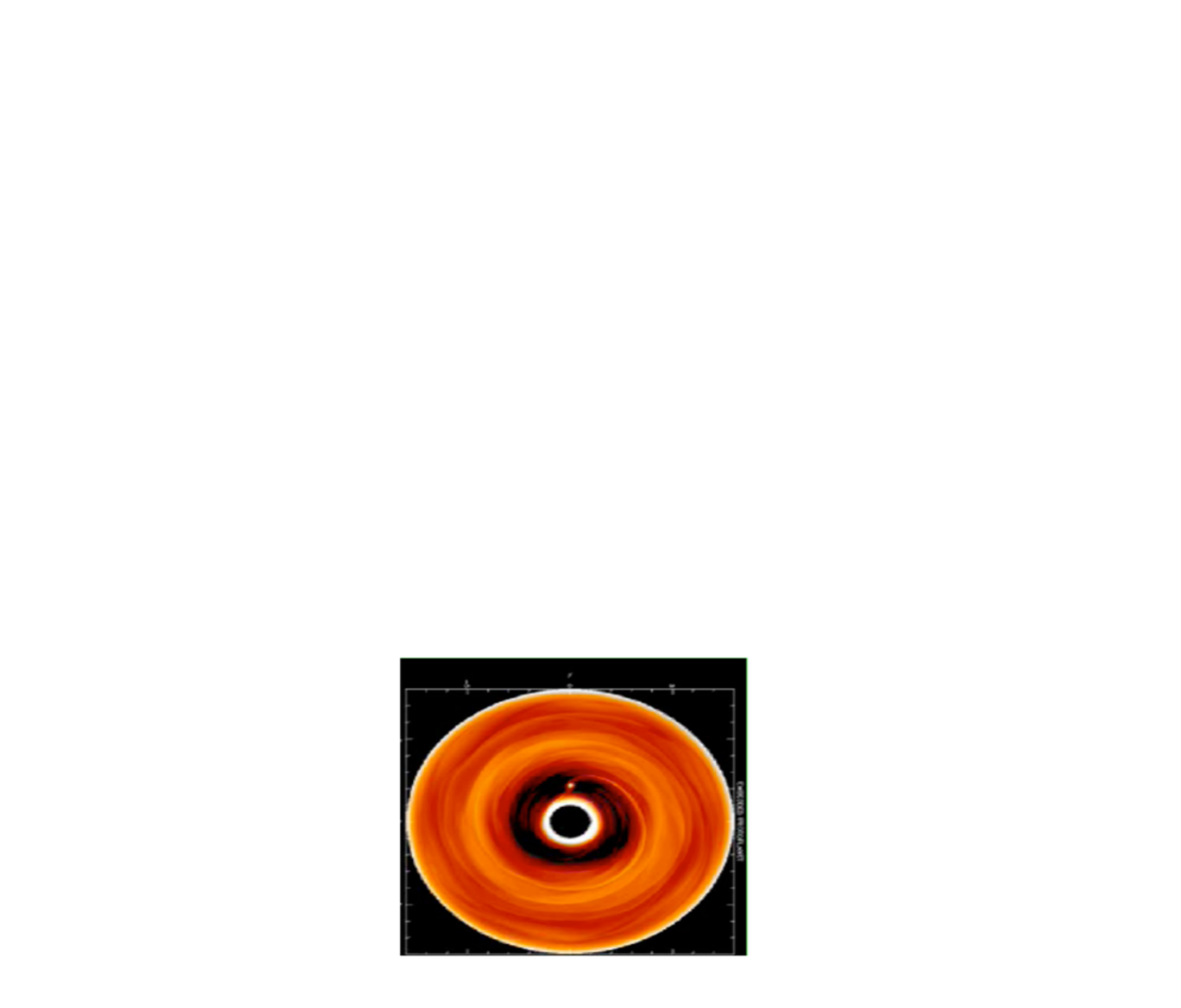}
\caption{ 	Left panel: A five Earth mass planet orbits in the type I migration regime
	while embedded in a laminar
	disk with a surface density profile 
	similar to that of  the minimum mass solar nebula (MMSN). The form of the surface density is indicated (light high dark low).
	Note the outer wake is stronger leading to a negative torque on the planet.
	Middle panel:  A schematic illustration of the part of the horseshoe region in the neighbourhood of a low mass planet.
	The direction of  flow along streamlines is indicated. Note that the streamlines that leave the plot at the bottom
	wrap around to become the streamlines entering the plot from the top.
	Right panel:  A five Jupiter mass orbiting in a prominent gap in a disk with MHD turbulence
	is illustrated  in the  right  panel. Note that  the direction of orbital rotation is anticlockwise.}
\label{Fig1}
\end{figure*}
	\section{Disc planet interaction and the main types of migration}\label{Gen}
	A planet of mass, $M_p,$  orbiting within a protoplanetary disc interacts gravitationally with
	nearby disk gas \citep[][]{Betal2014}.
	 We shall assume the planet, moving in a circular Keplerian orbit  with angular velocity, $\Omega_p,$ at radius, $r_p,$
	is moving hypersonically with Mach number, ${\cal M},$ in the range $10-50.$  
	The disk aspect ratio, or ratio of vertical semi-thickness to radius is $\sim 1/{\cal M}.$
         Consider nearby material that is streaming
         past the planet supersonically on  circular orbits  with the  minimum distance from the planet exceeding
         $2r_p/(3{\cal M}).$ Assuming that pressure effects can be neglected we can view  this material
         as undergoing a scattering as it passes by \citep[see eg.][]{LP1979, LP1993, PT2006}. 
	\subsection{Scattering Calculation}
	Consider  a ring of disk material orbiting interior to the planet
	at radius $r_p - b,$ where $b$ is small.
	As a fluid element of the ring   undergoes  its closest approach,
	its trajectory is  deflected through  an angle  $\delta$ given by
	\begin{equation}
	\tan(\delta/2) = {GM_p}/({b v^2}),
	\end{equation}
	where the relative speed $v= 3\Omega_pb/2.$
	For small $\delta$  the specific angular momentum transferred to the planet
	is 
	\begin{equation}
	\Delta j =r_pv(1-\cos\delta) \sim {r_pv\delta^2}/{2} = 2r_pv\left({GM_p}/{(bv^2)}\right)^2
	\end{equation}
         Considering the interior disk to consist of an ensemble of rings
          the rate of angular momentum  transfer to the planet is
         \begin{equation}
        { dJ}/{dt}= \int^{\infty}_{b_0}\Sigma \left( {3\Omega_p b}/{2}\right) \Delta j db= 
         ({8}/{27})\Sigma r_p^4 q^2\Omega_p^2\left( {r_p}/{b_0}\right)^3, \label{Jdot}
         \end {equation} 
          where $\Sigma$ is the dic surface density, $b_0,$ is the smallest allowed $b$ which may correspond to
          an inner disk edge and $q=M_p/ M_*.$ 	A corresponding calculation  and expression
          applies to material orbiting exterior to the planet with the transfer of angular momentum
           being from the planet to the disc which rotates more slowly.
           
           For an embedded low mass planet the cut off distance  $b_0$ should be the distance at which the flow becomes supersonic,     
           thus $b_0=2r_p/(3{\cal M}).$ For large mass planets that make a gap in the disc,
           $b_0$ should be the distance to the gap edge. 
            The net torque on the
           planet  then depends on the difference between  contributions from the inner and outer disc. 

	\subsection{The effect of coorbital material}\label{com}
	Coorbital material no longer streams past the planet but  undergoes horseshoe
	turns, reversing direction as it approaches the planet (eg. as in going from A to A' in  Fig. \ref{Fig1} (middle panel).
	As such a turn  moves disc material from  exterior  to  interior to the planet, angular momentum is transferred to it.
	After moving around the horseshoe material  that was at A'  undergoes a close turn from  from C to C' .
	Consequently angular momentum is removed from the planet. It then  moves round the horseshoe  back to A.
	For a non zero net torque there must be a difference between the two  types of closely approaching material.
	This has to be brought about by rapid enough transport of angular momentum and possibly heat brought about by  material
	 outside  the coorbital region.
	If this does not occur 
	the torque  saturates  as the two types of approach cancel each other out.
	The unsaturated torque on low mass  planets is found to depend on the gradients of  the inverse specific vorticity, $\Sigma/\Omega_p,$
         and the entropy 
         if the material is not  barotropic or efficiently cooled. 
         A well known example  is the planet trap \citep[][]{Metal2006} where the surface density, and hence $\Sigma / \Omega_p$  decreases rapidly inwards.
         Then angular momentum transport associated with the horseshoe turn $A \rightarrow A'$ will exceed that
         from the  turn $C \rightarrow C'$ in Fig. \ref{Fig1}  resulting in net transport to the planet that can halt inward migration.
         Note that  coorbital dynamics 
         is also affected by
         entropy gradients \citep{PM2006},  magnetic fields \citep[eg.][]{T2003, Getal2013} and heat transport between the
         planet and coorbital material \citep{M2017}.

	\subsection{Type I migration}\label{T1m}
	This affects  planets  with masses up to those characteristic of  ice giants  embedded in a protoplanetary disc.
	To estimate the  migration  rate we use equation (\ref{Jdot}) with,  $b_0= 2r_p/(3{\cal M}).$ 
	 The difference between outer and inner torques
	 introduces a factor  $\sim 5/{\cal M},$ which measures the departure from strict symmetry, 
	 with the torque from the outer disc being larger (see Fig.\ref{Fig1})
         Thus the  total torque acting on the planet is estimated as
	  \begin{equation}
        { dJ}/{dt}\sim 
         -({8}/{27})\Sigma r_p^4 q^2\Omega_p^2\left({r_p}/{b_0}\right)^3\left({5}/{{\cal M}}\right) =
         - 5\Sigma r_p^4 q^2\Omega_p^2 {\cal M}^2
           , \label{Jdot1}
         \end {equation}  
         leading to the  the inward migration rate 
           \begin{equation}
         \tau_{mig}  \sim  10^5 y. \left( {M_*
         (4\pi\Sigma r_p^2)^{-1}(500)^{-1}}\right) \left({2\pi\Omega_p^{-1}{ (12y.)^{-1}}}\right)\left({3\times 10^{-5}{q^{-1}}}\right)\left( 400{\cal M}^{-2}\right)
           .\label{Jdot2}
         \end {equation}  
        The effect of coorbital material should be included but  this is ineffective  in locally isothermal discs
	in which $\Sigma/\Omega_p$ is constant such the  MMSN.  
	The estimate (\ref{Jdot2}) then 
	 indicates a  rapid inward migration time scale $\sim 10^5 y.$ 
	for a $10M_{\oplus}$ planet  at   5 astronomical units $(au).$ This is much shorter than protoplanetary
	disc lifetimes $\sim 10^7 y.$,  leading some to arbitrarily drop planetary migration from consideration \citep[eg.][]{HM2012}.
	However, this neglects  coorbital torques in more general disc models  that could  reverse
	or halt  this migration  ( e.g. see discussion of a planet trap in Section. \ref{com}). 	
	Thus instead of   migrating inwards  protoplanets stall at  preferred locations 
	where mass growth may occur.

	   \subsection{Type II migration}\label{T2m}
	This applies to protoplanets massive enough 
	to make a gap in the disc.
	Typically,  their mass should exceed that of Saturn.
	\citep[for further discussion see the review by][]{PT2006}.
	The surface density profile adjusts so that angular momentum exchange with the planet, which  repels  disc material,
	is balanced by internal transport through  the disc at  gap edges. Denoting the angular momentum flow through the disc
	by, ${\cal F},$  making use of equation (\ref{Jdot})  this balance requires that at an edge 
	   \begin{equation}
        {\cal F}= 
         ({8}/{27})\Sigma r_p^4 q^2\Omega_p^2\left({r_p}/{b_0}\right)^3, \label{Jbal}
         \end {equation} 
	The cut off distance is the semi-width of the gap. We estimate a minimum value, $b_0 = (3/2)r_H= (3/2)(q/3)^{1/3}r_p,$
	where $r_H$ is the Hill radius.
	 For any realisable gap with larger $b_0$ we have
	   \begin{equation}
        {\cal{F}}\le
         0.26 \Sigma r_p^4 q\Omega_p^2, \label{Jbal1}.
         \end {equation} 
          For a disc in which angular momentum transport results from an effective kinematic viscosity, $\nu,$ 
          assumed to result from a form of turbulence,  ${\cal F}= 3\pi\nu\Sigma r_p^2\Omega_p.$ From
         (\ref{Jbal1}) we then get
             \begin{equation}
        {36\nu}/{( r_p^2\Omega_p})  \le q .\label{Jbal2}
         \end {equation} 
         This is marginally satisfied for a Saturn mass planet for a typical value of $\nu.$
         For a  {\it local}  gap to be produced we also require the planets gravity to dominate
         pressure effects within the Hill sphere leading to the requirement that $r_H /r_p= (q/3)^{1/3} >(1/{\cal M} )$   
          \citep[see eg.][]{LP1993}.
            A  five Jupiter mass orbiting in a prominent gap in a turbulent disk
	is   illustrated. in Fig.\ref{Fig1}.
          When there is a deep gap, even though there may be mass flow across it,
           torque balance across the gap results in radial migration  on the same time scale as the 
	general  local evolution of the disk as indicated by \citet{LP1986} (
	for more recent discussion of this aspect see \citet{Retal2018} and \citet{Setal2020}).
	This is typically $2\times 10^5 y$
	in standard models resembling the MMSN
	with $h = ({\cal M })^{-1} = 0.05$ at 5  $au$  \citep[see eg.][]{Netal2000}.
	However, this time increases with the local orbital period.
	 
	 \subsection{Type III Migration}\label{Type3}
          This  very rapid migration  can occur for planets that  form partial gaps in 
          sufficiently massive discs \citep{MP2003, Petal2008}. It involves  a feedback process operating through coorbital 
          torques that  greatly amplifies  the effect of an already existing  possibly  weak torque, $-{\cal T.}$ 
          
          Suppose the planet  migrates inwards at a rate ${\dot a}$ ( a corresponding parallel discussion applies for  outward migration).
           Matter flows through the horseshoe  region via the separatrix  at a rate ${\dot m} =2\pi\Sigma r_p |{\dot a}|.$
           This produces a torque acting on the planet of magnitude $|{\dot J}| = 2\pi\Sigma |{\dot a}|r_p^2\Omega_px_s,$
           where $x_s$ is the half width of the horseshoe region.  As the planet moves it  drags  coorbital material
           with it. Noting that the mass in this region is  $4\pi \Sigma_s r_p x_s,$ where $\Sigma_s$ is the surface density
           in the gap.
           This is added to the planet mass when considering torque balance which gives
           \begin{equation}
          \frac{1}{2} (M_p   +   4\pi \Sigma_s r_p x_s ) r_p\Omega_p |{\dot a}| =|{\dot J}| - {\cal T} = 2\pi\Sigma |{\dot a}|r_p^2\Omega_px_s - {\cal T}.\hspace{3mm}{\rm Hence}
          \hspace{3mm}  \frac{1}{2}  (M_p   - \delta m ) r_p\Omega_p |{\dot a}| = - {\cal T}.
           \end{equation}
           where,  $\delta m=  4\pi \Sigma_s r_p x_s (\Sigma -\Sigma_s), $ is the coorbital mass deficit. 
           When the coorbital mass
           deficit approaches the planet mass migration can increase rapidly, the direction determined by the seeding torque $-{\cal T}.$ 
           In practice the migration rate can increase until the time to migrate through the horseshoe region,  $2x_s/|{\dot a}|,$ becomes comparable to
           the horseshoe libration period $8\pi r_p/(3\Omega_px_s) .$ This leads to a characteristic fast migration time
           \begin {equation}
           r_p/|{\dot a}|= (2/3)(r_p/x_s)^2P_{orb}=( 8/3^{7/3})q^{-2/3}P_{orb},\label{T3}
          \end{equation}
            where we adopt $x_s=3r_H/2$  
           and $P_{orb}$ is the orbital period.
   \subsection{High eccentricity migration}
  After the protoplanetary disc has dispersed secular
   gravitational interactions amongst the planets in a multi-planetary system  may lead to a planet on an orbit with very high eccentricity
   which has close approaches to the central star leading to tidal interaction that circularises the orbit which conserves its angular momentum
    \citep[eg.][]{Wu2011, Naozetal2013, P2020}. 
   The semi-major axis may thus undergo a large decrease
   leading to an orbit with a period of a few days. High orbital inclinations relative to the initial orbital plane may also be generated. 
   In principle the Kozai-Lidov  mechanism could attain a similar result through the action of a binary companion in a highly inclined orbit.
  However, it has been argued that there are not enough companions of the required type for this to be viable 
   except in a small number of cases \citep[see][for a review]{DJ2018} .

	


\section{Important questions/goals for future studies}\label{goals} 
\subsection{Planet migration  in discs with MHD winds and ordered magnetic fields}\label{MigGP}
\begin{figure*}
\vspace{-1cm}
\hspace{-1cm} \includegraphics[trim=-1cm -0.7cm  0cm -0cm,clip,width=0.7\textwidth,angle=0]{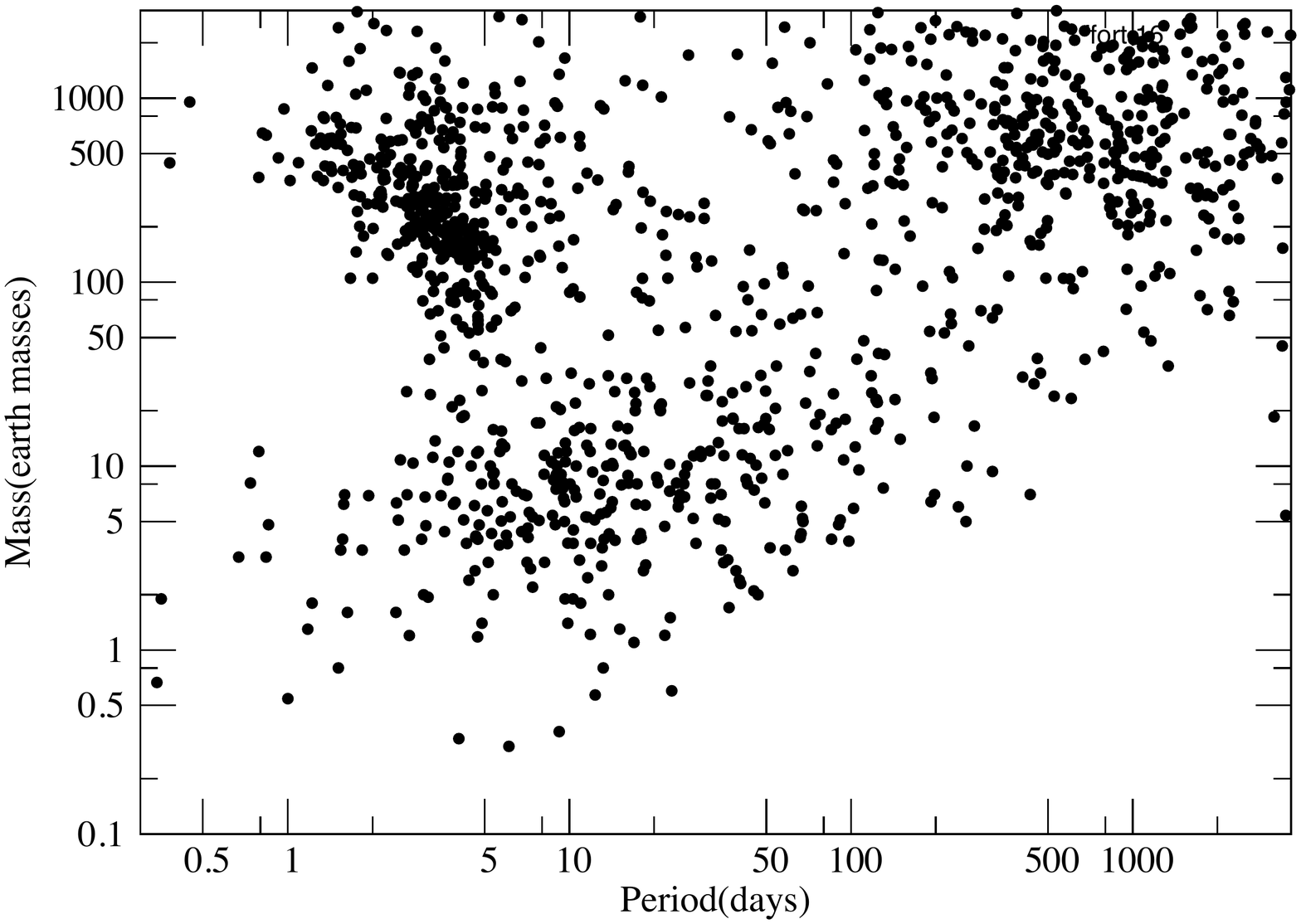}
\hspace{-0.8cm} \includegraphics[trim=0cm 0cm 6 0,clip,width=0.4\textwidth]{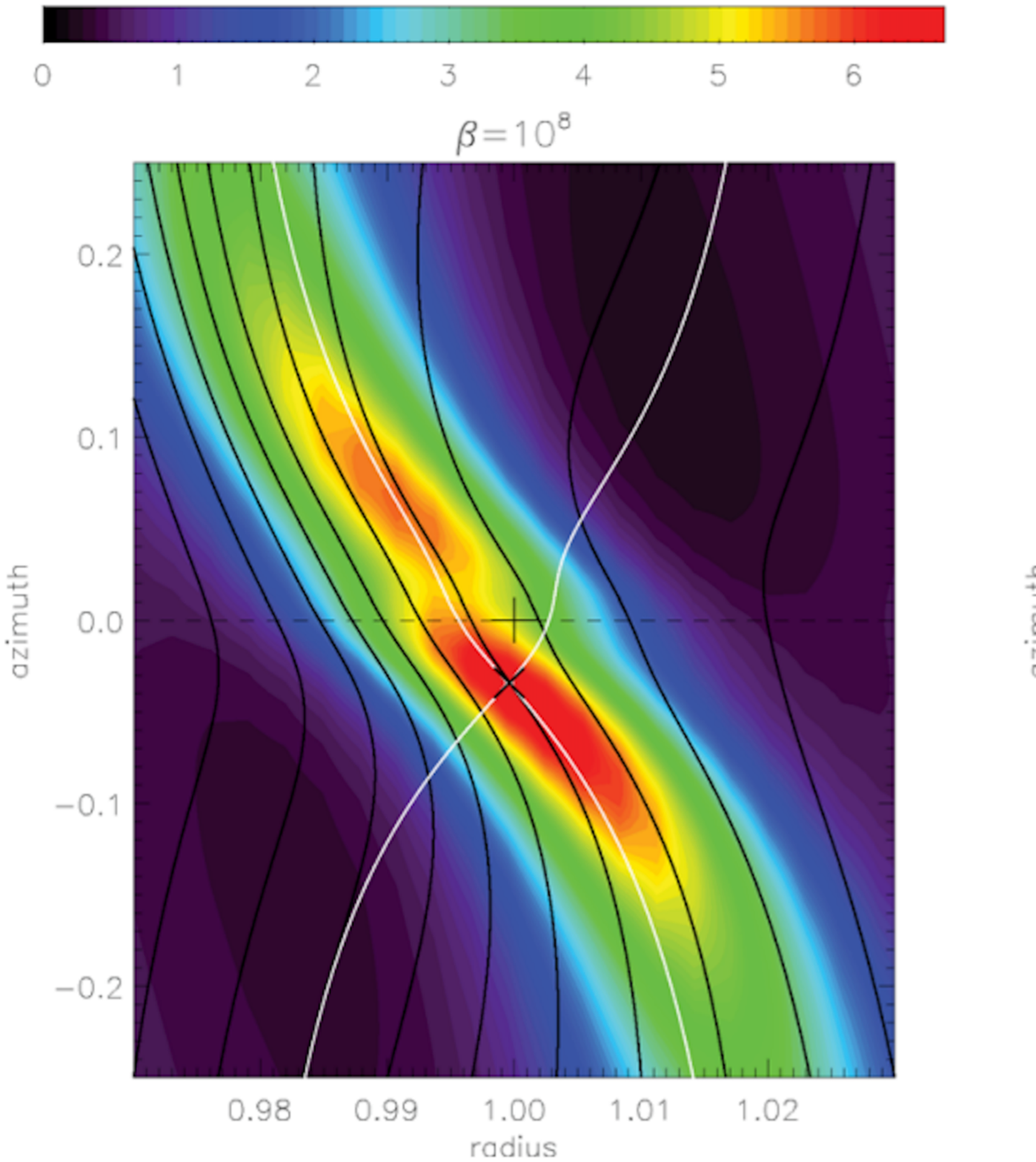}
\caption{ Left panel:  The dependence of the planet mass, $M_p,$ on orbital period,
	the data   was taken from 
	the NASA exoplanet archive: https://exoplanetarchive.ipac.caltech.edu/index.html (as  of January 2019).
	 Right panel: Magnetic energy contours in the coorbital region of a  $\sim 7M_{\oplus}$  planet, with location indicated by a + sign,
	   where there was an initial toroidal field.
	 The gas flow around the horseshoe turns produces regions of field concentration. The black lines denote field lines 
	 and the. white lines horseshoe separatrices (taken from \citet{Getal2013} ).}
\label{Fig3}
\end{figure*}

             Hot Jupiters have masses $ > 0.3 M_J$ and orbital periods $< 10 d.$ They can be seen as a group in the top left
             of the left panel of Fig. \ref{Fig3}.
	 About one in ten giant planet systems contain a Hot Jupiter \citep[see eg.][and references therein]{P2020}.
	 Disc migration  has been proposed  as a mechanism for moving them
	  from beyond the ice line  to their present close in orbits.  For a review of  the applicability of this  mechanism  as well as others involving gravitational 
	  interactions amongst  planets or with binary companions  leading to high eccentricity migration after the disc has dispersed see \citet{DJ2018}. 
	 Note that  the left panel of  Fig.\ref{Fig3} indicates a separation between giant planets with orbital periods $> \sim 100d$
	 that may have formed beyond an ice line with later modest migration
	 and the Hot Jupiters that underwent a more extreme form. Type II migration is the most likely mechanism
         applying to giant planets \citep[see eg.][]{LP1986, Betal2014}  with Hot Jupiters potentially undergoing more extreme type III migration
         \citep[see eg.][]{MP2003,Petal2007}. 
         
         As the time scale for type II migration at a few $au$ is significantly less than the disc lifetime
          \citep[eg.][]{Netal2000} a slowing down mechanism is needed. 
         Advantage may be taken of the scaling of the migration rate with viscosity when there are
          regions of the disc where the viscosity  is  expected to be  small.  This requires consideration of recent models that include
         magnetic fields and  non-ideal  magnetohydrodynamics (MHD)  
           \citep[eg.][]{B2017,Retal.2020}.
          Three dimensional (3D) simulations show  the disc inner region  to be  mainly laminar with large scale toroidal fields and  wind-driven accretion. 
          This is in contrast to earlier simulations that  studied gap formation assuming ideal MHD
           \citep[eg. the simulation in][illustrated in the right panel of Fig.1]{NP2003}.
           
           In the presence of MHD winds with negligible explicit viscosity, gap formation
           and type II  migration are still expected but with  the angular momentum loss rate
           from the outer disc required to supply the wind used in equation (\ref{Jbal1}).
            No such balance applies to the inner disc if it is disconnected
           from the outer disc. However, magnetic linkage  could affect this, enabling
           angular momentum transfer between the inner and outer disc independent of the planet. This may cause its
           migration to differ from that found for a viscous disc, potentially  slowing it down.  
           Simulations also need to be consider the role 
           of the vertical shear instability  \citep[eg.][]{SK2015}  and incorporate processes leading to photo-evaporation and disc dispersal
           in order to investigate final planetary configurations  \citep[eg.][]{AP2012}. 
           
           Studies of type III migration also need to be done with recent disc models in 3D.
           Here  there is significant material in the coorbital region, some of which may 
           be associated with vortex production that can interrupt  the migration \citep{LP2010}  as well as  accrete onto the planet.
           It has been assumed that the planet does not accrete significantly \citep[eg.][]{Petal2008},          
           though  significant accretion might inhibit the mechanism. 
            It is possible that the planet fills its critical Hill sphere or Roche lobe and is unable to accept  mass
           on a rapid type III migration time scale.
           We note  that the thermal time  scale of a planet filling its Roche lobe may be estimated as
           $t_{KH} \sim GM_p^2/(4\pi (2r_H/3)^3\sigma T_{e}^4),$  where $T_e$ is the effective temperature and $\sigma$ is Stefan's constant.
           This may be written
           \begin{equation} 
            t_{KH}\sim 81\pi M_p/(8P_{orb}^2\sigma T_{e}^{4}) \sim 1255(M_p/M_{\saturn}) (P_{orb}/(1y.))^{-3}(T_e/(300K))^{-4}P_{orb} ,
          \end{equation}
          which may plausibly exceed the characteristic time for type III migration given by (\ref{T3}) leading to  thermally restricted accretion
          for a Saturn mass planet at $\sim 1au.$
           The issues highlighted here  need to be investigated in more detail. 
           \subsection{Tidal dissipation and high eccentricity migration}
            Orbital evolution produced as a result of tidal interaction with the central star may be important for giant planets brought close in by disc migration
           or high eccentricity migration occurring after the disc has dispersed. In the latter case, as the energy that has to be dissipated may greatly exceed 
           the binding energy of the planet \citep{PBI2004}, the interaction may be strongly nonlinear, potentially
           leading to  disruption \citep{Guill et al.2011}. In this context the issue of how much energy is dissipated in the
           planet  and whether it can be efficiently radiated away  is important for determining  possible final orbital parameters 
           of hot Jupiters. Further studies are needed.
       
           \subsection{Coorbital torques, saturation and  halting type I migration}
           The role of coorbital torques in halting  type I migration has been mentioned in Section \ref{T1m}.
             As the action of coorbital torques or horseshoe drag may determine disc locations where planets can grow
             it is important that they be investigated with the most accurate disc models that incorporate magnetic fields
             where relevant. These can prevent the torques from saturating when there is an effective viscosity (see Section \ref{com}) as shown
             for a disc with ideal MHD and a simple equation of state by \citet{Betal2011}.
              However, what happens in the non ideal case with ordered fields and winds is unclear.
                         
             The  effect of  a large scale toroidal magnetic field has been considered in two dimensional (2D) model discs 
             by \citet{T2003} for ideal MHD and \citet{Getal2013}
             when there is enough magnetic diffusivity to enable disc gas to diffuse across field lines as horse shoe turns are executed
             ( see the right panel of Fig. \ref{Fig3}). In the former inviscid case, horseshoe turns are prevented through the operation
             of Alfv{\`e}n resonances and the migration can reverse. In the latter case 
             field concentrations can produce surface density changes that  modify the torques   potentially reversing  migration ( see Fig. \ref{Fig3}).
              These simulations adopted a laminar viscosity with Renolds' number  of  $1.2\times 10^5$
              and  Prandtl number  of unity in most cases. This needs to be looked at in 3D models with non ideal MHD without
              applied  viscosity. One might anticipate  instabilities leading to
              vortex production  for sufficiently large planet masses \citep[eg.][]{Letal2009}. 
              
              Another  issue associated with horseshoe turns in 3D models 
              is the role of vertical stratification \citep[see][]{Metal2020}. They find that buoyancy oscillations lead to a speeding up
              inward migration. This needs to be considered  with active MHD and thermal diffusion.
              Additional effects arise from heat transport connecting the planet and disc material \citep{M2017}.
              Asymmetries in the horseshoe  region (note the displacement of the planet from the
              central stagnation point in the right panel of Fig.\ref{Fig3}) and an offset between the circular motion of the
              planet and disc material  lead to asymmetries in the heat transport and density
              structure that  can produce migration stalling torques. The effects of accretion of mass and angular momemntum
              and a magnetic connection should be studied in 3D.
           
              \subsection{Multiplanet systems, commensurabilities and conversion from inward to outward migration, convergent to divergent migration
              and the Grand Tack}\label{Multi}
               Systems of planets often do not behave as if the components were independent.
               Convergent migration of planet pairs  leads to mean motion resonances yet their incidence 
               is modest, being  about 30\% of giant planet pairs  with period ratios  $<3.5$ and 20\% of all pairs with period ratio $3$ are close to
               a commensurability \citep[see eg.][for discussion]{P2020}. There are rare systems of  three or more. planets forming resonant chains.
                If  convergent migration  of  components is  common  during or post
                formation  some disruptive  mechanism is required to disrupt them either before or after the disc  dispersal,
                 \citep[eg.][]{Ietal2017}. Up to now studies of disc planet interactions in these contexts have mostly adopted simple
               disc models and been in 2D.  
               
               The joint migration of Jupiter and Saturn was first studied by \citet{M2001}.
               They found that the  system switched from inward to outward migration after an initial rapid inward migration of Saturn
               caused the system to lock into a 3:2 commensurability.  This scenario has been adopted in the Grand Tack scenario proposed to operate
               during the  formation of the solar system \citep[see eg.][]{Petal2014}.
               The process which occurs when an inwardly migrating giant planet is caught up by a lighter one  has been proposed as a mechanism
               that can push planets out to large distances  \citep[see][]{Cetal2009}.
               These simulations involve type II migration as well as fast  possibly type III migration of an outer lighter giant planet. Thus the issues
               highlighted in Section \ref{MigGP} apply. Consideration of  realistic disc models with allowance for the  mass growth 
                fed by material in  coorbital regions  is needed to see if the outer planet  always has the lowest mass as required in these scenarios.   
               
               \subsubsection{Formation and subsequent evolution of commensurabilites}
               Giant planets with periods $>100d$ may not have migrated  through large
               distances (see left panel  of Fig.\ref{Fig3}).
                The absence of super Earths with bare  cores composed of volatiles
               and the relative rarity  of  commensurabiities in two planet systems 
               indicates the same  applies to them \citep[see eg.][]{PT2019,P2020}.
                But the  existence  of near commensurabilites 
                supports a picture of at least restricted migration in some cases.
               The formation of low order commensurabilites through the  convergent migration of two planets is well documented
               \citep[see eg.][]{Betal2014}. However,  initially convergent migration can be reversed through wake planet interactions
               when  the density response produced by  one planet reacts back on the other. This has been  seen when giant planets
               interact with incoming super Earths \citep[][]{Petal2012} and for pairs of Neptune mass planets \citep[][]{B2013}. It results in a departure
               from commensurability and may help reduce the number of expected strict commensurabilities.
               These  studies  should be extended to incorporate the final stages 
               of the evolution of the protoplanetary disk. The location and evolution of structural features
               constituting  migration traps and interaction with planetesimals and dust will  be important.
               \subsection*{Summary}
               Studies of disc planet interactions began with simple models  before the discovery of exoplanets about forty years ago.
               While studies of migration resulting from disc free processes such as gravitational interactions leading to high eccentricity migration
               followed by  tidal interaction with the central star began shortly after the first discoveries of exoplanets.
               Advances in our understanding of protoplanetary discs coupled with improving  numerical tools and computational resources
               offer the prospect of more realistic modelling  of planet formation and early evolution.  This will lead to improved 
               comparisons with observations of structured protoplanetary discs \citep[see eg.][]{Francis2020} as well as 
               improved predictions for the form of planetary systems.
\bibliography{sample63}{}

\begin{thebibliography}{100}
\bibitem[Alexaander \& Pascucci(2012)]{AP2012} 
Alexander,  R., Pascucci,  I., 2012, MNRAS,  422, L82 
\bibitem[Bai(2017)]{B2017} 
         Bai, X.-N., 2017,  ApJ, 845,  75, 30
\bibitem[Baruteau et al.(2011)]{Betal2011}        
         Baruteau C., Fromang S., Nelson R. P., Masset F., 2011, A\&A, 533, A84
\bibitem[Baruteau \& Papaloizou( 2013)]{B2013}          
         Baruteau, C.,  Papaloizou, J., 2013, MNRAS, 430, 1764
 \bibitem[Baruteau et al.(2014)]{Betal2014}           
         Baruteau C., et al., 2014, Protostars and Planets VI, pp 667–689
\bibitem[Crida et al.(2009)]{Cetal2009}     
         Crida, A.,  Masset, F.,  Morbidelli, A., 2009, ApJL, 705, L148
\bibitem[Dawson \& Johnson(2018)]{DJ2018}       
         Dawson, R. I., Johnson, J. A.,  2018,  ARA\&A, 56, 175
\bibitem[Francis \& van der Marel(2020)]{Francis2020} 
        Francis, L.,  van der Marel, N.,  2020, ApJ, 892, 111,
\bibitem[Guilet et al.(2013)]{Getal2013}   
         Guilet, J, Baruteau, C., Papaloizou, J. C. B., (2013),  MNRAS, 430, 1764
\bibitem[Guillochon et al.(2011)]{Guill et al.2011}   
        Guillochon, J.,  Ramirez-Ruiz, E.,  Lin, D.,  2011, ApJ, 762, 37 
\bibitem[Hansen \& Murray(2012)]{HM2012}                  
        Hansen, B. M. S.,  Murray, N., 2012. ApJ, 158, 16
\bibitem[Ivanov \& Papaloizou(2004)]{PBI2004}  
       Ivanov, P.B., Papaloizou, J. C. B., 2004, MNRAS, 347, 437.
\bibitem[Izidoro et al.(2017)]{Ietal2017}
         Izidoro,  A., et al.,
          2017, MNRAS,  470, 1750         
\bibitem[Li et al.(2009)]{Letal2009}
         Li H., Lubow S. H., Li S., Lin D. N. C., 2009, ApJ, 690, L52
\bibitem[Lin \& Papaloizou(1979)]{LP1979} 
	Lin, D. N. C., Papaloizou, J. C. B., 1979, MNRAS, 186, 799 
\bibitem[Lin \& Papaloizou(1986)]{LP1986} 
	Lin, D. N. C., Papaloizou, J. C. B., 1986, ApJ, 309, 846 
\bibitem[Lin \& Papaloizou(1993)]{LP1993} 
	Lin, D. N. C., Papaloizou, J. C. B., 1993, Protostars and Planets III, 
	Univ. of Arizona Press, Tucson, AZ, 749
\bibitem[Lin \& Papaloizou(2010)]{LP2010} 	
	Lin M.-K., Papaloizou J. C. B., 2010, MNRAS, 405, 1473
\bibitem[Masset(2017)]{M2017} 
	Masset, F.S., 2017, MNRAS, 472,  4204 
\bibitem[Masset et al.(2006)]{Metal2006} 
	Masset, F., S., Morbidelli, A., Crida, A., Ferreira, J., 2006, 
	ApJ, 642, 478
\bibitem[Masset \& Papaloizou(2003)]{MP2003} 
         Masset, F., Papaloizou, J. C. B., 2003, ApJ, 588, 494
\bibitem[Masset \& Snellgrove(2001)]{M2001} 
	Masset, F., Snellgrove, M., 2001. MNRAS, 320, L55
\bibitem[Mcnally et al.(2020)]{Metal2020} 
        McNally, C. P., et al.,
        2020, MNRAS, 493, 4382
\bibitem[Naoz et al.(2013)]{Naozetal2013} 
        Naoz, S.,  et al., 2013, MNRAS, 431, 2155
\bibitem[Nelson(2005)]{N2005} 
        Nelson R. P., 2005, A\&A, 443, 1067
\bibitem[Nelson et al.(2000)]{Netal2000} 
        Nelson, R. P.,  Papaloizou, J. C. B., Masset, F.,  Kley, W., 2000, MNRAS, 318, 18
\bibitem[Nelson \& Papaloizou(2003)]{NP2003} 
        Nelson, R. P.,  Papaloizou, J. C. B., 2003, MNRAS, 339, 993
\bibitem[Papaloizou(2020)]{P2020}  
        Papaloizou, J., 2020,  The orbital architecture of exoplanetary systems [1].
        
      In Oxford Research Encyclopaedia of Planetary Science, Oxford University Press,
        
      doi:https://doi.org/10.1093/acrefore/9780190647926.013.190
\bibitem[Papaloizou et al.(2007)]{Petal2007} 
      Papaloizou J. C. B., 
       Nelson R. P., Kley W., Masset F. S., Artymowicz P.,
       2007, Protostars and Planets V, Univ. of Arizona Press, Tucson, AZ, 655
\bibitem[Papaloizou \& Terquem(2006)]{PT2006} 
         Papaloizou J. C. B., Terquem C., 2006, Reports on Progress in Physics, 69, 119
\bibitem[Papaloizou \& Terquem(2019)]{PT2019} 
        Papaloizou, J.C.B., Terquem, C., 2019. MNRAS, 482, 530
\bibitem[Paardekooper \& Mellema(2006)]{PM2006}     
        Paardekooper S. J., Mellema G., 2006, A\&A, 459, L17
\bibitem[Peplinski et al.(2008)]{Petal2008}   
        Pepliński, A., Artymowicz, P.,  Mellema, G. 2008, MNRAS, 386, 179
\bibitem[Pierens et al.(2014)]{Petal2014} 
       Pierens, A., Raymond, S. N., Nesvorny, D., Morbidelli, A., 2014,
       ApJL, 795. L11
\bibitem[Podlewska-Gaca, et al.(2012)]{Petal2012}
        Podlewska-Gaca, E., Papaloizou, J. C. B., Szuszkiewicz, E., 2012, MNRAS, 421, 1736
\bibitem[Rein(2012)]{R2012}     
        Rein, H., 2012,MNRAS, 427, L21.
\bibitem[Riols et al.(2020)]{Retal.2020}    
        Riols, A., Lesur, G., Menard, F., 2020, A\&A, 639, A95
\bibitem[Robert et al.(2018)]{Retal2018} 
	Robert, C. M. T., Crida, A., Lega, E., M{\' e}heut H., et al., 2018, A\&A, 617, A98 
\bibitem[Scardoni et al.(2020)]{Setal2020} 
	Scardoni, C. E.,  Rosotti, G. P.,  Lodato, G.,  Clarke, C. J., 2020, 
	MNRAS, 492, 1319
\bibitem[Stoll \& Kley(2015)]{SK2015} 
	Stoll, M. H. R.,  Kley, W., 2015, A\&A, 572, A77
\bibitem[Terquem(2003)]{T2003} 	
	Terquem, C. E. J .M. L. J., 2003, MNRAS, 341, 1157
\bibitem[Wu(2011)]{Wu2011} 	
        Wu, Y., Lithwick, Y., 2011, ApJ, 735, 109
\end{thebibliography}
\bibliographystyle{aasjournal}


\end{document}